\documentclass[sigplan,nonacm]{acmart}
\usepackage{graphicx} 
\usepackage{lipsum}
\usepackage{tabularray, color}

\usepackage[]{hyperref}
\title{Implementing and Optimizing the Scaled Dot-Product Attention on Streaming Dataflow}

\author{Gina Sohn}
\email{ginasohn@stanford.edu}
\affiliation{%
  \institution{Stanford University}
  \city{Stanford}
  \country{USA}
}

\author{Nathan Zhang}
\email{stanfurd@stanford.edu}
\affiliation{%
  \institution{Stanford University}
  \city{Stanford}
  \country{USA}
}

\author{Kunle Olukotun}
\email{kunle@stanford.edu}
\affiliation{%
  \institution{Stanford University}
  \city{Stanford}
  \country{USA}
}

\date{April 2024}

\begin{document}

\begin{abstract}
    Transformer models serve as the backbone of many state-of-the-art language models, and most use the scaled dot-product attention (SDPA) mechanism to capture relationships between tokens. However, the straightforward implementation of SDPA has quadratic compute and memory complexity with respect to the sequence length.
    On processor architectures such as GPUs and TPUs, there is a robust body of prior work. However, little work has been performed on non-processor architectures.
In this work, we show how the architecture and execution model of Streaming Dataflow Accelerators can help tackle this challenge. We first define abstract hardware that adopts a streaming execution model, and we implement a cycle-accurate simulator of the abstract hardware using the Dataflow Abstract Machine simulation framework.
Second, we implement the naive SDPA algorithm on this abstract hardware and show it requires linear $(O(N))$ intermediate memory.
Third, we then modify the naive algorithm, taking inspiration from prior processor-oriented works, by reordering the multiplication and division operations. 
Finally, we map the modified algorithm to abstract hardware, and confirm that the implementation computes SDPA at full throughput while only using a constant amount $(O(1))$ of intermediate memory.
\end{abstract}
    
\maketitle 

\section{Introduction}
Transformer models~\cite{attention2017} are widely used for various language, audio, and vision tasks. Over time, researchers have discovered that long sequence lengths are essential to capturing long-range dependencies and process high-resolution images~\cite{brown2020language, chowdhery2022palm, openai2023gpt4}. However, scaled dot-product attention (SDPA), the core operation in most transformer models, has quadratic memory complexity with respect to the sequence length. There has been a wide range of work to tackle this challenge, such as approximate attention~\cite{kitaev2020reformer,katharopoulos2020transformers,beltagy2020longformer}, using alternative mixing mechanisms~\cite{fu2024monarch, fu2022hungry}, and optimizations to reduce reads and writes between different levels in the memory hierarchy~\cite{flashattention2022, dao2023flashattention}. However, these works almost uniformly targeted processor-like architectures such as CPUs, TPUs, and GPUs. 

\begin{figure}
    \centering
    \includegraphics[width=0.65\columnwidth]{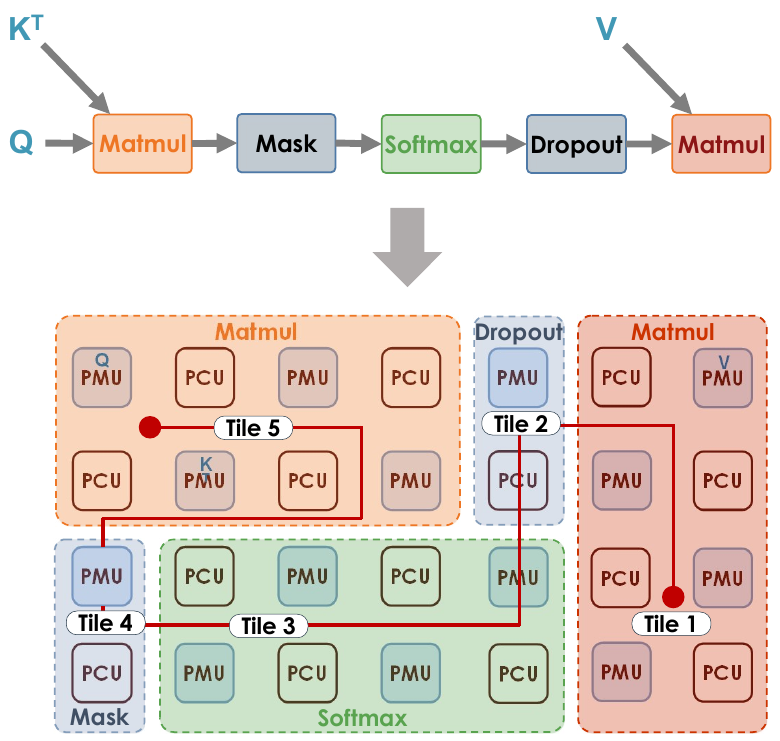}
    \caption{An abstract diagram of the architecture and execution model for streaming dataflow accelerators}
    \label{fig:exec-model}
\end{figure}

In this work, we will instead implement and optimize the attention algorithms on streaming dataflow accelerators. Streaming dataflow accelerators are array architectures of reconfigurable compute and memory units. As shown in Figure ~\ref{fig:exec-model}, its execution model spatially maps the operations in the computation graph to hardware units by configuring the compute and memory units accordingly. It then pipelines the execution between different operations. Some examples of streaming dataflow accelerators include Coarse-grained Reconfigurable Architectures~\cite{10.1145/3534933,carsello2022amber} and Reconfigurable Dataflow Architectures~\cite{prabhakar2017plasticine, rucker2023revet}. The streaming dataflow paradigm provides several performance benefits. First, it enables exploiting a high degree of parallelism due to dataflow execution and deeply pipelined execution. Second, it requires less memory footprint and bandwidth as operation fusion reduces off-chip memory accesses due to intermediate data. 

To elucidate how the streaming execution model can reduce the memory complexity of the attention algorithm, we first define an abstract streaming dataflow accelerator based on Parallel Patterns~\cite{parallelpatterns}. We then map the attention algorithm to the abstract hardware and show that the implementation achieves linear memory complexity for intermediate data. Lastly, we apply algorithmic changes inspired by prior work on other architectures, and provide an implementation of SDPA that requires only a constant amount of intermediate memory.

\definecolor{Alto}{rgb}{0.87,0.87,0.87}
\begingroup
\setlength{\tabcolsep}{1.5pt}
\begin{table*}
\small
\centering
\begin{tblr}{
  row{1} = {Alto},
  hlines,
  vlines,
}
Node & Behavior\\
Map (f: func) & Applies the function \textbf{f} to every element in the input stream\\
Reduce (n: Int) (init: T) (f: func) & {Reduces across \textbf{n} element in the input stream using the given function.\\The output will be enqueued to the output stream after \textbf{n} elements are reduced.\\ \textbf{init} is the initial value of the accumulator.}\\
MemReduce (n: Int) (init: Mem[T]) (f: func) & {Similar to Reduce, but executes higher order reduction on memory elements \\instead of scalar values.}\\
Repeat (n: Int) & Repeats every scalar in the input stream \textbf{n} times.\\
Scan (n: Int) (init: T) (updt: func) (f: func) & {On every new input element, update the state using the \textbf{updt} function.\\The state is initialized to the \textbf{init} value after scanning every \textbf{n} elements.\\On every input element, apply the function \textbf{f} and dequeue to the output stream.}
\end{tblr}
\caption{The nodes in our abstract hardware based on Parallel Patterns.}
\label{table:abstract-hw}
\end{table*}
\endgroup

\section{The Abstract Streaming Dataflow Hardware}

Prior work has shown that Parallel Patterns can simplify the generation process of optimized configurations for configurable hardware~\cite{parallelpatterns, prabhakar2017plasticine, Koeplinger2018Spatial}.
Our abstract hardware consists of nodes such as \textbf{Map}, \textbf{Reduce}, and \textbf{Scan} (Table~\ref{table:abstract-hw}). Each node can be further lowered to a configuration of the physical compute and memory units.
To verify the functional correctness of the implementation and confirm it uses a linear amount of intermediate memory without any performance loss, we develop a cycle-accurate simulator for each node in the abstract hardware using the Dataflow Abstract Machine simulation framework \cite{DAM}.

\begin{figure*}
    \centering
    \includegraphics[width=0.85\linewidth]{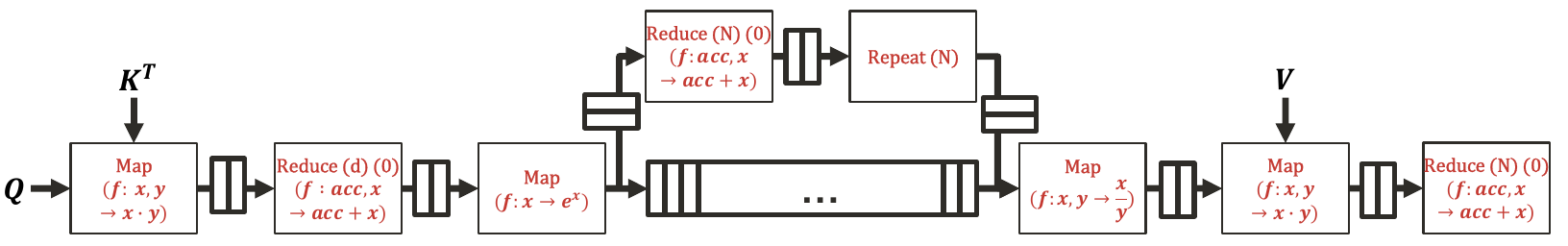}
    \caption{Implementation of attention using Parallel Patterns. The depth of the short FIFOs is set to 2, and the depth of the Long FIFO is set to $N+2$. Each node can be mapped to a configuration of a set of compute and memory units in a streaming dataflow hardware.}
    \label{fig:naive-attn}
\end{figure*}

\begin{figure*}
    \centering
    \includegraphics[width=0.9\linewidth]{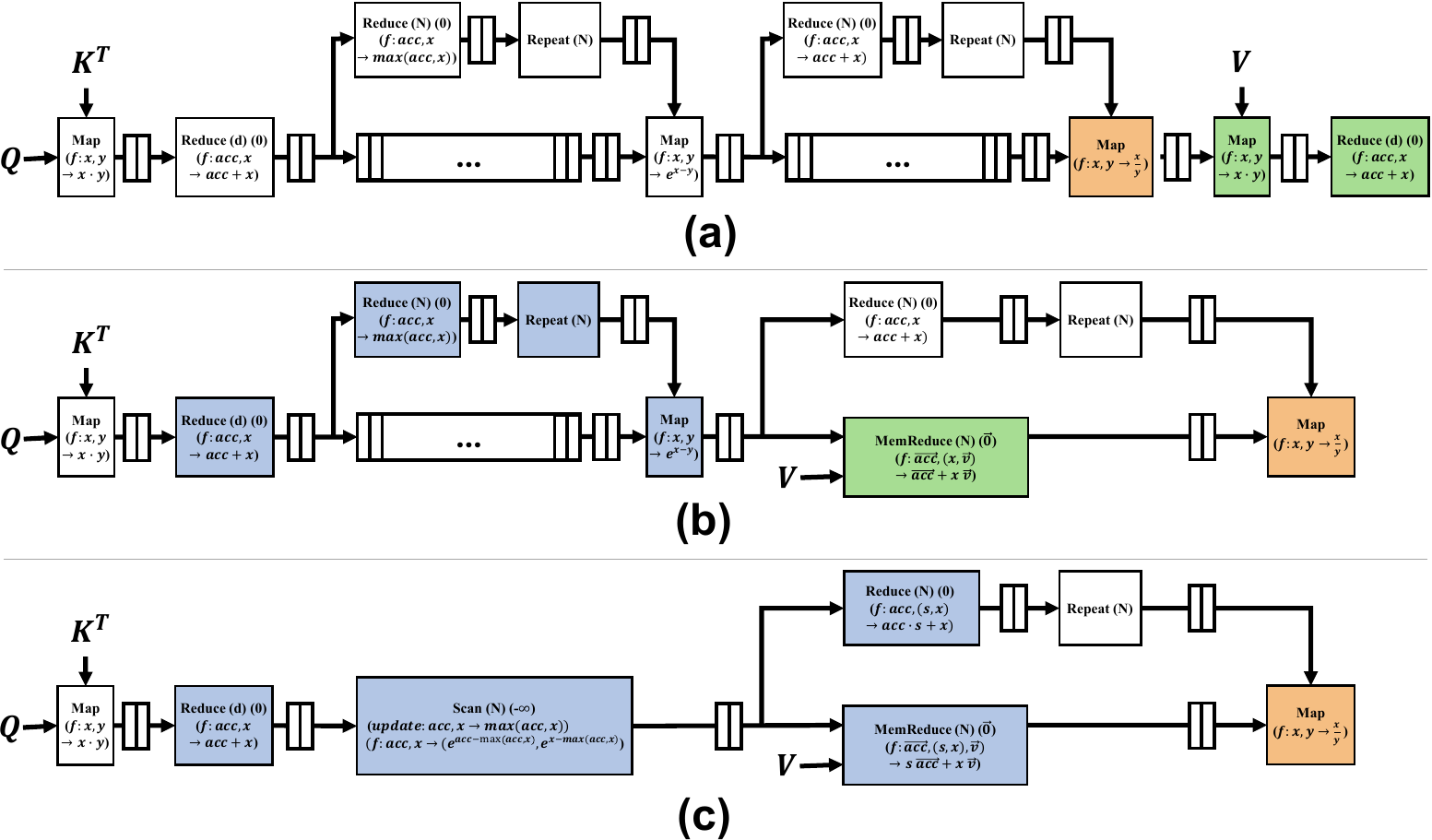}
    \caption{Implementations of algorithms using Parallel Patterns. (a) The attention algorithm using softmax with scaling. (b) The attention algorithm with reordered division. (c) The memory-free-attention algorithm.}
    \label{fig:mem-free-attn}
\end{figure*}

\section{Standard Attention Implementation}
Inside the Transformer model, three matrices $Q, K,$ and $V$ are generated based on the input sequence and weights. Given that there are $N$ tokens in the input sequence where each token is a $d$-wide vector, the shape of $Q, K,$ and $V$ is ${N\times d}$, and the attention algorithm can be expressed as follows:
\begin{equation} \label{eq:standard-attn}
S = QK^T, \hspace{8pt} P = softmax(S) = \frac{e^{s_{ij}}}{\sum_j e^{s_{ij}}}, \hspace{8pt} O = PV 
\end{equation}

The algorithm produces $N^2$-sized intermediate matrices $S$ and $P$, leading to a time and memory complexity quadratic to the sequence length. 
However, on a streaming execution model, it is possible to implement the same algorithm to have a sub-quadratic, in other words, asymptotically lower, memory complexity without any algorithmic changes. This starts from the observation that the attention algorithm can be decomposed into row-wise operations. 
Given that $\vec{q_i}$ is the i-th row of the $Q$ matrix, Equation \ref{eq:standard-attn} can be decomposed into row-wise operations as follows ($\vec{s_i}, \vec{p_i}, \vec{o_i}$ are the i-th row of matrix $S, P, O$ respectively) :

\begin{equation} \label{eq:row-standard-attn}
\vec{s_i} = \vec{q_i}K^T, \hspace{8pt} \vec{p_i} = softmax(\vec{s_i}) = \frac{e^{s_{ij}}}{\sum_j e^{s_{ij}}}, \hspace{8pt} \vec{o_i} = \vec{p_i}V 
\end{equation}

The rows of matrix $Q$ can be streamed into compute units that execute the operations in Equation \ref{eq:row-standard-attn}, and the computation between different rows can be pipelined. This has the effect of fusing operations to compute each row of the output matrix $O$. Therefore, the memory complexity for intermediate data will reduce to the required intermediate data to compute a \emph{single} row.

The mapping of the attention algorithm to our abstract hardware can be found in Figure~\ref{fig:naive-attn}. We configure each node as shown in the figure and set the depth of the short FIFOs to two and the long FIFO to $N+2$. We compare this with the baseline, where all the FIFOs are set to have infinite depth (this will be the peak throughput scenario). We confirmed that the implementation in Figure ~\ref{fig:naive-attn} only requires $O(N)$ intermediate memory while running in full throughput. More detailed experiment results can be found in the case study in the Dataflow Abstract Machine work~\cite{DAM}.

\section{Memory-free Attention Implementation}
By modifying the algorithm, we can further obtain an implementation that requires a constant amount ($=O(1)$) of intermediate memory. 
In the previous section, the reason why we need a $O(N)$-long FIFO is that there are two paths from the second \textbf{Map} unit to the third \textbf{Map} unit with different latencies. The path that contains a row-wise reduction produces the output only after processing the last element in the row, and the following \textbf{Map} unit performs an element-wise operation on the pair of inputs from each path. The other path needs a deep FIFO that can contain the outputs from the second \textbf{Map} unit until the reduction path produces its output to avoid deadlock. Therefore, when the consumer performs an element-wise operation with the inputs from each path, the two divergent paths should have matched latency to avoid introducing this long FIFO.

This can be done by (1) reordering the division operation in softmax with the following matrix multiplication based on the distributive law, (2) using the running sum instead of the row-wise sum during the softmax operation, and (3) rescaling by the difference between the old running sum and latest running sum during the scalar reduction in softmax and the vector reduction in the matrix multiplication. For this implementation, we use softmax with scaling \cite{flashattention2022, kitaev2020reformer,milakov2018online,rabe2021n2memory}, which is widely used in practice for better numerical stability. The modified algorithm can be described as:
\begin{equation}\label{eq:mem-free-qkt}
s_{ij} = \sum_k q_{ik}\cdot k_{kj} 
\end{equation}

Given $\forall j<0\;\; m_{ij}= -\infty$, for all $i > 0$:
\begin{equation}\label{eq:mem-free-scan}
\begin{aligned}
m_{ij} = max(m_{i(j-1)}, s_{ij}) \\
\Delta_{ij} = e^{m_{i(j-1)}-m_{ij}} \\
e_{ij} = e^{s_{ij}-m_{ij}}
\end{aligned}\end{equation}

Given $\forall j<0\;\; r_{ij} = 0$ and $\vec{l_{ij}} = \vec{0}$, for all $i > 0$:

\begin{equation}\label{eq:mem-free-reduce}
\begin{aligned}
r_{ij} = r_{i(j-1)}\cdot \Delta_{ij} + e_{ij} \\
\vec{l_{ij}} = \vec{l}_{i(j-1)}\cdot \Delta_{ij} + e_{ij}\cdot \vec{v_j}
\end{aligned}\end{equation}

\begin{equation}\label{eq:mem-free-div}
    \vec{o_i} = \frac{\vec{v}_{iN}}{r_{iN}}
\end{equation}


In the attention algorithm using softmax with scaling (Figure~\ref{fig:mem-free-attn}(a)), there are two pairs of divergent paths with unbalanced latency due to the row-wise reduction. The second pair of unbalanced paths are handled by reordering the division operation in softmax and the matrix multiplication. As shown in Figure~\ref{fig:mem-free-attn}(b), by making the matrix multiplication's reduction happen in parallel with the row-wise summation, it balances the latency of both paths and eliminates the $O(N)$-long FIFO.

The first pair of unbalanced paths remaining in Figure~\ref{fig:mem-free-attn}(b) can be eliminated using a running max instead of the row-wise max. As the following operations are reductions, the difference between the latest running max and the running max at the point when each element was accumulated can be adjusted by rescaling the accumulated values. This, in turn, eliminates the deep FIFO by converting the reduction operation into an element-wise scan operation.

The implementation in Figure~\ref{fig:mem-free-attn}(c) is simulated by configuring each node in our simulator accordingly and setting the depth of the short FIFOs to two. This is compared with the baseline, where all the FIFOs are set to have infinite depth (this will be the peak throughput scenario). We confirmed that this implementation only requires $O(1)$ intermediate memory to run in full throughput. More detailed experiment results can be found in the case study in the Dataflow Abstract Machine paper~\cite{DAM}.

\appendix

\bibliographystyle{plain}
\bibliography{references}

\end{document}